\documentclass[twocolumn, showpacs, pre, amsmath, amssymb, superscriptaddress]{revtex4}

\usepackage{graphicx}
\usepackage{dcolumn}
\usepackage{bm}
\usepackage{latexsym}
\usepackage{times}

\begin{document}


\title{Module identification in bipartite and directed networks}

\author{Roger Guimer\`a}

\affiliation{Northwestern Institute on Complex Systems (NICO) and
Department of Chemical and Biological Engineering, Northwestern
University, Evanston, IL 60208, USA}

\author{Marta Sales-Pardo}

\affiliation{Northwestern Institute on Complex Systems (NICO) and
Department of Chemical and Biological Engineering, Northwestern
University, Evanston, IL 60208, USA}

\author{Lu\'{\i}s A. Nunes Amaral}

\affiliation{Northwestern Institute on Complex Systems (NICO) and
Department of Chemical and Biological Engineering, Northwestern
University, Evanston, IL 60208, USA}


\begin{abstract}
  Modularity is one of the most prominent properties of real-world
  complex networks. Here, we address the issue of module
  identification in two important classes of networks: bipartite
  networks and directed unipartite networks. Nodes in bipartite
  networks are divided into two non-overlapping sets, and the links
  must have one end node from each set. Directed unipartite networks
  only have one type of nodes, but links have an origin and an end. We
  show that directed unipartite networks can be conviniently
  represented as bipartite networks for module identification
  purposes. We report a novel approach especially suited for module
  detection in bipartite networks, and define a set of random networks
  that enable us to validate the new approach.
\end{abstract}


\date{\today}

\maketitle


Units in physical, chemical, biological, technological, and social
systems interact with each other defining complex networks that are
neither fully regular nor fully
random \cite{albert02,newman03,amaral04}. Among the most prominent and
ubiquitous properties of these networks is their modular
structure \cite{girvan02, newman03}, that is, the existence of
distinct groups of nodes with an excess of connections to each other
and fewer connections to other nodes in the network.

The existence of modular structure is important in several
regards. First, modules critically affect the dynamic behavior of the
system. The modular structure of the air transportation system
\cite{guimera05b}, for example, is likely to slow down the spread of
viruses at an international scale \cite{colizza06a} and thus somewhat
minimize the effects of high-connectivity nodes that may otherwise
function as ``super- spreaders''
\cite{pastor-satorras01a,liljeros03}. Second, different modules in a
complex modular network can have different structural properties
\cite{guimera07}. Therefore, characterizing the network using only
global average properties may result in the misrepresentation of the
structure of many, if not all, of the modules. Finally, the modular
structure of networks is likely responsible for at least some of the
correlations (e.g. degree-degree correlations
\cite{newman02,pastor-satorras01b,maslov02,maslov04,colizza06}), that
have attracted the interest of researchers in recent years
\cite{guimera07}.

For the above reasons, considerable attention has been given to the
development of algorithms and theoretical frameworks to identify and
quantify the modular structure of networks (see \cite{danon05} and
references therein). However, current research activity has paid
little attention, except for a few studies in sociology
\cite{borgatti97,doreian04}, to the problem of identifying modules in
a special and important class of networks known as bipartite networks
(or graphs). Nodes in bipartite networks are divided into two
non-overlapping sets, and the links must have one end node from each
set. Examples of systems that are more suitably represented as
bipartite networks include:

\begin{itemize}

\item Protein-protein interaction
networks \cite{uetz00,jeong01,maslov02,li04} obtained from yeast two
hybrid screening: one set of nodes represents the {\it bait} proteins
and the other set represents the {\it prey} or {\it library}
proteins. Two proteins, a bait and a library protein, are connected if
the library protein binds to the bait.

\item Plant-animal mutualistic networks \cite{jordano87,bascompte03}:
one set represents animal species and the other set represents plant
species. Links indicate mutualistic relationships between animals and
plants (for example, a certain bird species feeding on a plant species
and dispersing its seeds).

\item Scientific publication networks
\cite{newman01b,borner04,guimera05c}: one set represents scientists
and the other set represents publications. A link between a scientist
and a publication indicates that the scientist is one of the authors
of the publication.

\item Artistic collaboration networks
\cite{gleiser03,uzzi05,guimera05c}: one set represents artists and the
other teams. A link indicates the participation of an artist in a
team.

\end{itemize}

Another important class of networks for which no sound module
identification methods are available are unipartite directed
networks. Examples of directed unipartite networks include:

\begin{itemize}

\item Food webs \cite{williams00,stouffer05}: nodes represent species
and links indicate trophic interactions in an ecosystem.

\item Gene regulatory networks \cite{barabasi04}: nodes are genes and
links indicate regulatory interactions.

\end{itemize}

The usual approach to identify modules in directed networks is to
disregard the directionality of the connections, which will fail when
different modules are defined based on incoming and outgoing links.

Here, we address the issue of module identification in complex
bipartite networks. We start by reviewing the approaches that are
currently used heuristically and aprioristically to solve this
problem. We then suggest a new approach especially suited for module
detection in bipartite networks, and define a set of random networks
that permit the evaluation of the accuracy of the different
approaches. We then discuss how it is possible to use the same
formalism to identify modules in directed unipartite networks. Our
method enables one to independently identify groups of nodes with
similar outgoing connections and groups of nodes with similar incoming
connections.

\section{Background}

For simplicity, from now on we denote the two sets of nodes in the
bipartite network as the set of {\it actors} and the set of {\it
teams}, respectively. Given a bipartite network, we are interested in
identifying groups (modules) of actors that are closely connected to
each other through co-participation in many teams. Of course, one is
free to select which set of nodes in a given network is the ``actor
set'' and which one is the ``team set,'' so one can identify modules
in either or both set of nodes.

We require any module-identification algorithm to fulfill two quite
general conditions: (i) the algorithm needs to be {\it network
independent}; and (ii) given the list of links in the network, the
algorithm must determine not only a good partition of the nodes into
modules, but also the {\it number of modules and their sizes}.

The first condition is somewhat trivial. We just make it explicit to
exclude algorithms that are designed to work with a particular network
or family of networks, but that will otherwise fail with broad
families of networks (for example, large networks or sparse/dense
networks).

The second condition is much more substantial, as it makes clear the
difference between the module-identification problem and the graph
partitioning problem in computer science, in which both the number of
groups and the sizes of the groups are fixed. To use a unipartite
network analogy, given a set of 120 people attending a wedding and
information about who knows whom, the graph partitioning problem is
analogous to optimally setting 12 tables with 10 people in each
table. In contrast, the module-identification problem is analogous to
identifying ``natural'' groups of people, for example the different
families or distinct groups of friends.

The second condition also excludes algorithms (based, for example, on
hierarchical clustering or principal component analysis
\cite{everitt01}) that project network data into some low-dimensional
space without specifying the location of the boundaries separating the
groups. For example, given a dendogram generated using hierarchical
clustering, one still needs to decide where to ``cut it'' in order to
obtain the relevant modules. To be sure, one can propose a combination
of algorithms that first project the data into some low-dimensional
space and then set the boundaries, and assess the accuracy of the
method. In general, however, one cannot {\it evaluate the performance
of hierarchical clustering}, given that hierarchical clustering does
not provide a single solution to module-identification
problem. Neither can one test the infinite combinations of
dimensionality reduction algorithms with techniques for the actual
selection of modules.

Freeman \cite{freeman03} has recently compiled a collection of 21
algorithms that have been used in the social networks literature to
identify modules in bipartite networks. To the best of our
understanding none of the algorithms described there satisfies the two
conditions above. Among the statistical physics community, on the
other hand, the common practice is to project the bipartite network
onto a unipartite actors' network, and then identify modules in the
projection. In the scientists' projection of a scientific publication
network, for example, two scientists are connected if they have
coauthored one or more papers. The caveat of this approach is that,
even if the projection is weighted (by for example, the number of
papers coauthored by a pair of scientists), some information of the
original bipartite network, like the sizes of the teams, is lost in
the projection. Here, we suggest an alternative to existing approaches
to identify modules in complex bipartite networks.

\section{Modularity for bipartite networks}
\label{s-modularity}

A widely used and quite successful method for the identification of
modules in unipartite networks is the maximization of a modularity
function. Although this method has limitations
\cite{fortunato07,sales-pardo??,fortunato07a}, it yields the most
accurate results reported in the literature for a wide family of
random networks with prescribed modular structure
\cite{guimera05a,guimera05,danon05}.

In the same spirit, here we define a modularity function that, upon
optimization, yields a partition of the actors in a bipartite network
into modules. By doing this, the module identification problem becomes
a combinatorial optimization problem that is analogous to the
identification of the ground state of a disordered magnetic system
\cite{guimera04,reichardt06}.

A ubiquitous modularity function for unipartite networks is the
Newman-Girvan modularity \cite{newman04b}. The rationale behind this
modularity is that, in a modular network, links are not homogeneously
distributed. Thus, a partition with high modularity is such that the
density of links inside modules is significantly higher than the
random expectation for such density. Specifically, the modularity
$\mathcal{M(P)}$ of a partition $\mathcal{P}$ of a network into
modules is
\begin{equation}
\mathcal{M(P)} = \sum_{s=1}^{N_M} \left[ \frac{l_{s}}{L} - \left(
\frac{d_s}{2L} \right)^2 \right] \, ,
\label{e-modularity}
\end{equation}
where $N_M$ is the number of modules, $L$ is the number of links in
the network, $l_{s}$ is the number of links between nodes in module
$s$, and $d_s$ is the sum of the degrees of the nodes in module
$s$. Then $l_s/L$ is the fraction of links inside module $s$, and
$(d_s/2L)^2$ is an approximation (assuming that self-links and
multiple links between nodes are allowed) to the fraction of links one
would expect to have inside the module from chance alone.

We define a new modularity $\mathcal{M_B(P)}$ that can be applied to
identify modules in bipartite networks. We start by considering the
expected number of times that actor $i$ belongs to a team comprised of
$m_a$ actors:
\begin{equation}
m_a \frac{t_i}{\sum_k t_k}\; ,
\label{e-1}
\end{equation}
where $t_i$ is the total number of teams to which actor $i$
belongs. Similarly, the expected number of times that two actors $i$
and $j$ belong to team $a$ is
\begin{equation}
m_a(m_a-1) \frac{t_i t_j}{\left( \sum_k t_k \right)^2}\;.
\label{e-2}
\end{equation}
Therefore, the average number of teams in which $i$ and $j$ are
expected to be together is
\begin{equation}
\frac{\sum_a m_a(m_a-1)}{\left( \sum_a m_a \right)^2} t_i t_j \; ,
\label{e-common}
\end{equation}
where we have used the identity $\sum_a m_a = \sum_k t_k$.  Note that
$\sum_a m_a(m_a-1)$ and $\left( \sum_a m_a \right)^2$ are global
network properties, which do not depend on the pair of actors
considered.

Equation (\ref{e-common}) enables us to define the bipartite
modularity as the cumulative deviation from the random expectation
\begin{equation}
  \mathcal{M_B(P)} = \sum_{s=1}^{N_M} \left[ \frac{\sum_{i \ne j \in
      s} c_{ij}}{\sum_a m_a(m_a -1)} - \frac{\sum_{i \ne j \in s} t_i
      t_j}{\left( \sum_a m_a \right)^2} \right] \; ,
  \label{e-modularityb}
\end{equation}
where $c_{ij}$ is the actual number of teams in which $i$ and $j$ are
together. For convenience, we exclude the irrelevant diagonal term
$i=j$ from the sums \footnote{Including the diagonal term would only
shift the modularity by a constant, and is therefore irrelevant.}, and
normalize the modularity so that $\mathcal{M_B} \rightarrow 1$ when:
(i) all actors in each team belong to a single module ($\sum_s \sum_{i
\ne j \in s} c_{ij} = \sum_a m_a(m_a -1)$), and (ii) the random
expectation for pairs of nodes being in the same team is small
($\sum_s \sum_{i \ne j \in s} t_i t_j \ll \left( \sum_a m_a
\right)^2$).

As in the derivation of Eq.~(\ref{e-modularity}), the null model
implicit in Eqs.~(\ref{e-1}) and (\ref{e-2}) is such that one could,
in principle, have multiple connections between an actor and a
team. In most cases this situation would not make sense, so the null
model is only appropriate when $m_a$ and $t_i$ are much smaller than
$\sum_a m_a$, for all $a$ and all $i$.

\section{Model bipartite networks with modular structure}
\label{s-modelnets}

\begin{figure}[t!]
\centerline{
\includegraphics*[width=\columnwidth]{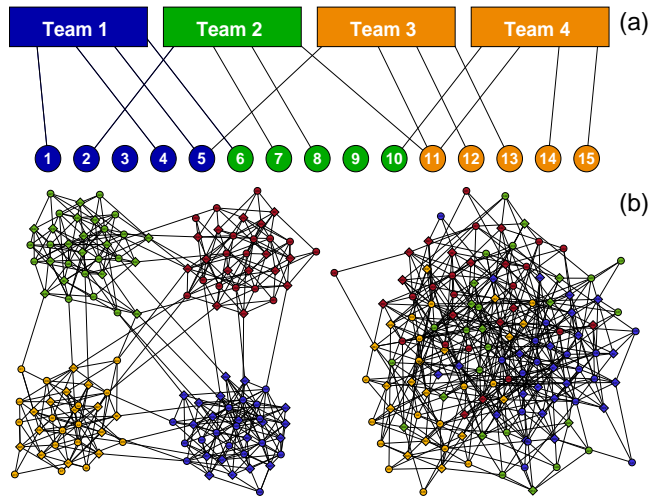}
}
\caption{
  Model random bipartite networks with modular structure.
  (a) Nodes are divided into two sets, actors (circles) and teams
  (rectangles). Each color represents a different module in the
  actors' set, and teams of a given color are more likely to contain
  actors of their color (see text).
  (b) Two sample networks with $N_M=4$ modules, with 16 actors
  (circles) each, and $N_T=64$ teams (diamonds), with $m=7$ actors
  each. The network on the left has a strong modular structure,
  $p=0.9$, while the modular structure is less well defined on the
  right, $p=0.5$ (see text for the definition of $p$).
}
\label{f-model}
\end{figure}

Ensembles of random networks with prescribed modular structure
\cite{girvan02} enable one to assess algorithm's performance
quantitatively, and thus to compare the performance of different
algorithms. Here, we introduce an ensemble of random bipartite
networks with prescribed modular structure (Fig.~\ref{f-model}).
%

We start by dividing the actors into $N_M$ of modules; each module $s$
comprises $S_s$ nodes. For clarity, we use different ``colors'' for
different modules. The network is then created assuming that actors
that belong to the same module have a higher probability of being
together in a team than actors that belong to different modules
\footnote{This is, to some extent, an implicit definition of what
modularity means in bipartite networks, in the same way that ``higher
linkage probability inside modules'' is a definition of what
modularity means in unipartite networks.}. Specifically, we proceed by
creating $N_T$ teams as follows:
\begin{itemize}

\item Create team $a$.

\item Select the number $m_a$ of actors in the team.

\item Select the {\it color} $c_a$ of the team, that is, the module
that will contribute, in principle, the most actors to the team.

\item For each spot in the team: (i) with probability $p$, select the
actor from the pool of actors that have the same color as the team;
(ii) otherwise, select an actor at random with equal probability. The
parameter $p$, which we call {\it team homogeneity}, thus quantifies
how homogeneous a team is. In the limiting cases, for $p=1$ all the
actors in the team belong to the same module and modules are perfectly
segregated, whereas for $p=0$ the color of the teams is irrelevant,
actors are perfectly mixed and the network does not have a modular
structure.

\end{itemize}

\section{Results}
\label{s-results}

We next investigate the performance of different module identification
algorithms in both model networks with predefined modular structure,
and in a simple real network that shows some interesting features.

We consider three approaches for the identification of modules in
bipartite networks. First, we consider the {\it unweighted projection}
(UWP) approach. Within this approach, we start by building the
projection of the bipartite network into the actors space. Then we
consider the projection as a regular unipartite network and use the
modularity given in Eq.~(\ref{e-modularity}).

Next, we consider the {\it weighted projection} (WP) approach. Within
this approach, we start by building the weighted projection of the
bipartite network. In the weighted projection, actors are connected if
they are together in one or more teams, and the weight $w_{ij}$ of the
link indicates the number of teams in which the two actors are
together (thus, $w_{ij} = c_{ij})$. We then use the simplest
generalization to weighted networks of the modularity in
Eq.~(\ref{e-modularity})
\begin{equation}
  \mathcal{M_W(P)} = \sum_{s=1}^{N_M} \left[ \frac{w^{\rm int}_{s}}{W}
    - \left( \frac{w^{\rm all}_s}{2W} \right)^2 \right] \, ,
  \label{e-modularityw}
\end{equation}
where $W=\sum_{i \ge j}w_{ij}$, $w^{\rm int}_{s}$ is the sum of the
weights of the links within module $s$, and $w^{\rm all}_{s} = \sum_{i
\in s} \sum_j w_{ij}$.

Finally, we consider the {\it bipartite} (B) approach. Within this
approach, we consider the whole bipartite network and use the
modularity introduced in Eq.~(\ref{e-modularityb}).

In all cases, we maximize the modularity using simulated annealing
\cite{kirkpatrick83}. Several alternatives have been suggested to
maximize the modularity including greedy search \cite{newman04a},
extremal optimization \cite{duch05}, and spectral methods
\cite{newman06,newman06b}. In general, there is a trade-off between
accuracy and execution time, with simulated annealing being the most
accurate method \cite{danon05}, but at present too slow to deal
properly with networks comprising hundreds of thousands or millions of
nodes.

\subsection{Model bipartite networks}

We consider the performance of the different module identification
approaches when applied to the model bipartite networks described
above. We assess the performance of an algorithm by comparing the
partitions it returns to the predefined group structure. Specifically,
we use the mutual information $I_{AB}$ \cite{danon05} between
partitions $A$ and $B$ to quantify the performance of the algorithms
\begin{equation}
  I_{AB} = \frac{-2 \sum_{i=1}^{N_M^A} \sum_{j=1}^{N_M^B} n_{ij}^{AB}
    \log \left( \frac{n_{ij}^{AB} S}{n_i^A n_j^B}\right)}
    {\sum_{i=1}^{N_M^A} n_i^A \log \left( \frac{n_i^A}{S} \right) +
    \sum_{j=1}^{N_M^B} n_j^B \log \left( \frac{n_j^B}{S} \right)}\,.
\end{equation}
Here, $S$ is the total number of nodes in the network, $N_{M}^{A}$ is
the number of modules in partition $A$, $n_{i}^{A}$ is the number of
nodes in module $i$ of partition $A$, and $n_{ij}^{AB}$ is the number
of nodes that are in module $i$ of partition $A$ and in module $j$ of
partition $B$. The mutual information between partitions A and B is 1
if both partitions are identical, and 0 if they are uncorrelated.

In the simplest version of the model all modules have the same number
of nodes, all teams have the same size, and the color of each team is
set assuming equal probability for each color. Unless otherwise
stated, we build networks with $N_M=4$ modules, each of them
comprising 32 actors, and $N_T=128$ teams of size $m=14$.

\subsubsection{Team homogeneity}

We first investigate how team homogeneity $p$ affects algorithm
performance. For $p=1$, all the actors in a team belong to the same
module, and any reasonable algorithm must perfectly identify the
modular structure of the network; thus $I=1$. Conversely, for $p=0$,
actors are perfectly mixed in teams, and all algorithms will return
random partitions due to small fluctuations \cite{guimera04}; thus
$I=0$. Any $p>0$ will provide a signal that an algorithm can, in
principle, extract.

\begin{figure}[t!]
  \centerline{
    \includegraphics*[width=.75\columnwidth]
		     {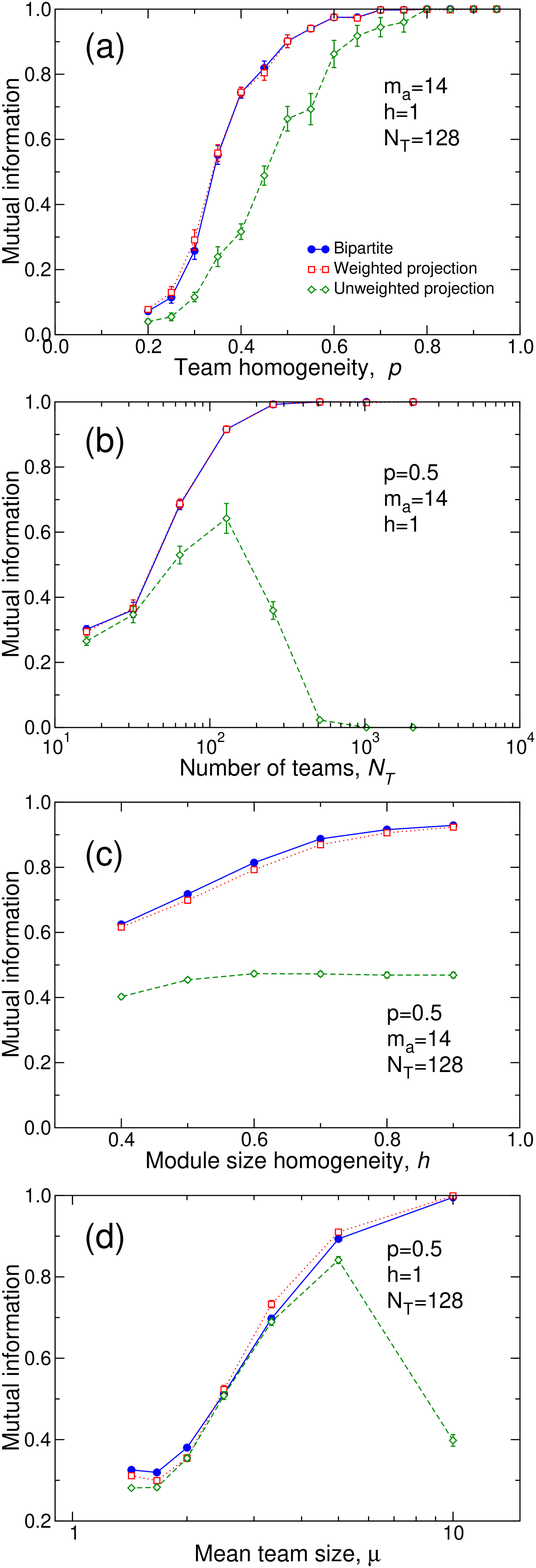}
  }
  \caption{Algorithm performance as a function of: (a) team
    homogeneity $p$ (simulation parameters: $N_M=4$, $S_s=32$ for all
    modules); (b) number of teams $N_T$ (simulation parameters:
    $N_M=4$, $S_s=32$ for all modules); (c) module size homogeneity
    $h$ (simulation parameters $N_M=6$, $132$ nodes); and (d) mean
    team size $\mu$ (simulation parameters: $N_M=4$, $S_s=32$ for all
    modules). Error bars indicate the standard error.}
  \label{f-performance}
\end{figure}
%
As shown in Fig.~\ref{f-performance}(a), the UWP approach performs
systematically and significantly worse than the weighted projection
and the bipartite algorithms for all values of $p$. For the choice of
parameters described above, the last two algorithms start to be able
to identify the modular structure of the network for $p \approx
0.35$. For $p \ge 0.5$, one already finds $I>0.9$. The WP and the B
approaches yield indistinguishable results.

\subsubsection{Number of teams and average team size}

Team homogeneity is not the only parameter affecting algorithm
performance. For example, the number of teams $N_T$ in the network
critically affects the amount of information available to an
algorithm. Interestingly, the number of teams affects in different
ways the UWP approach on the one hand, and the WP and B approaches on
the other; Fig.~\ref{f-performance}(b). For the WP and B algorithms,
the larger $N_T$, the larger the amount of information and therefore
the easier the problem becomes. Indeed, even for very small values of
$p$, the signal to noise ratio can become significantly greater than 1
if $N_T$ is large enough. On the contrary, as the number of teams
increases the UWP becomes denser and denser and eventually becomes a
fully connected graph, from which the algorithm cannot extract any
useful information. Once more, the performance of the WP and the B
approaches are indistinguishable.

\subsubsection{Module size heterogeneity}

In real networks, modules will have (sometimes dramatically) different
sizes~\cite{danon06}. Given the sizes of the modules in a network, and
assuming that they are ordered so that $S_1 \ge S_2 \ge \dots \ge
S_{N_M}$, we define $h$ as the ratio of sizes between consecutive
modules (with integer rounding)
\begin{equation}
h = \frac{S_{i+1}}{S_i}\,.
\end{equation}
Additionally, we select the color of the teams with probabilities
proportional to the size of the corresponding module, so that all
actors participate, on average, in the same number of teams.

As we show in Fig.~\ref{f-performance}(c), we again observe that the
WP and the B approach perform similarly, and clearly outperform the
UWP approach for all values of $h$.

\subsubsection{Team size distribution}

All the results so far suggest that the WP approach and the B approach
yield results that are indistinguishable from each other. We know,
however, that differences do exist between both. The distribution of
team sizes, in particular, is taken into account in the B approach but
disregarded in the WP approach, and ``teams'' with $m=1$ are totally
disregarded in projection-based approaches, but not in the B approach.

We thus investigate what is the effect of the team size distribution
on the performance of the algorithms. Instead of considering that all
teams have the same size $m$, we now consider a distribution $p(m)$ of
team sizes. In particular, we consider a (displaced) geometric
distribution
\begin{equation}
p(m) = \frac{1}{\mu} \left( 1 - \frac{1}{\mu} \right)^{m-1} \quad ,
\quad m \ge 1\,,
\end{equation}
which is the discrete counterpart of the exponential distribution. The
distribution has mean $\langle m \rangle = \mu$.

As we show in Fig.~\ref{f-performance}(d), some small differences seem
to appear between the WP approach and the B approach, although it is
difficult to establish conclusively if these differences are
significant or not.

In the light of this, we investigate in more depth the relationship
between the bipartite modularity in Eq.~(\ref{e-modularityb}) and the
weighted extension of the unipartite modularity in
Eq.~(\ref{e-modularity}). As we show in the Appendix, the bipartite
modularity actually reduces to the weighted unipartite modularity (up
to an irrelevant additive constant) when all teams in the bipartite
network have the same size.

This observation explains why the WP and the B approach differ when
teams have unequal sizes \footnote{Even in cases in which all teams
have the same size, small systematic discrepancies can occur between
the WP and the B approach, given that the simulated annealing used in
each case is different in some implementation details.}. Although our
results suggest that each approach outperforms the other in certain
cases, we believe that Eq.~(\ref{e-modularityb}) is, in general,
preferable because it explicitly takes into account the distribution
of team sizes, while the weighted projection does not.

\subsection{Southern women dataset}

During the 1930s, ethnographers Allison Davis, Elizabeth Stubbs Davis,
J. G. St. Clair Drake, Burleight B. Gardner, and Mary R. Gardner
collected data on social stratification in the town of Natchez,
Mississippi \cite{davis41,freeman03}. Part of their field work
consisted in collecting data on women's attendance to social events in
the town. The researchers later analyzed the resulting women-event
bipartite network in the light of other social and ethnographic
variables. Since then, the dataset has become a {\it de facto}
standard for discussing bipartite networks in the social sciences
\cite{freeman03}.

\begin{figure}[t!]
\centerline{ \includegraphics*[width=0.8\columnwidth]{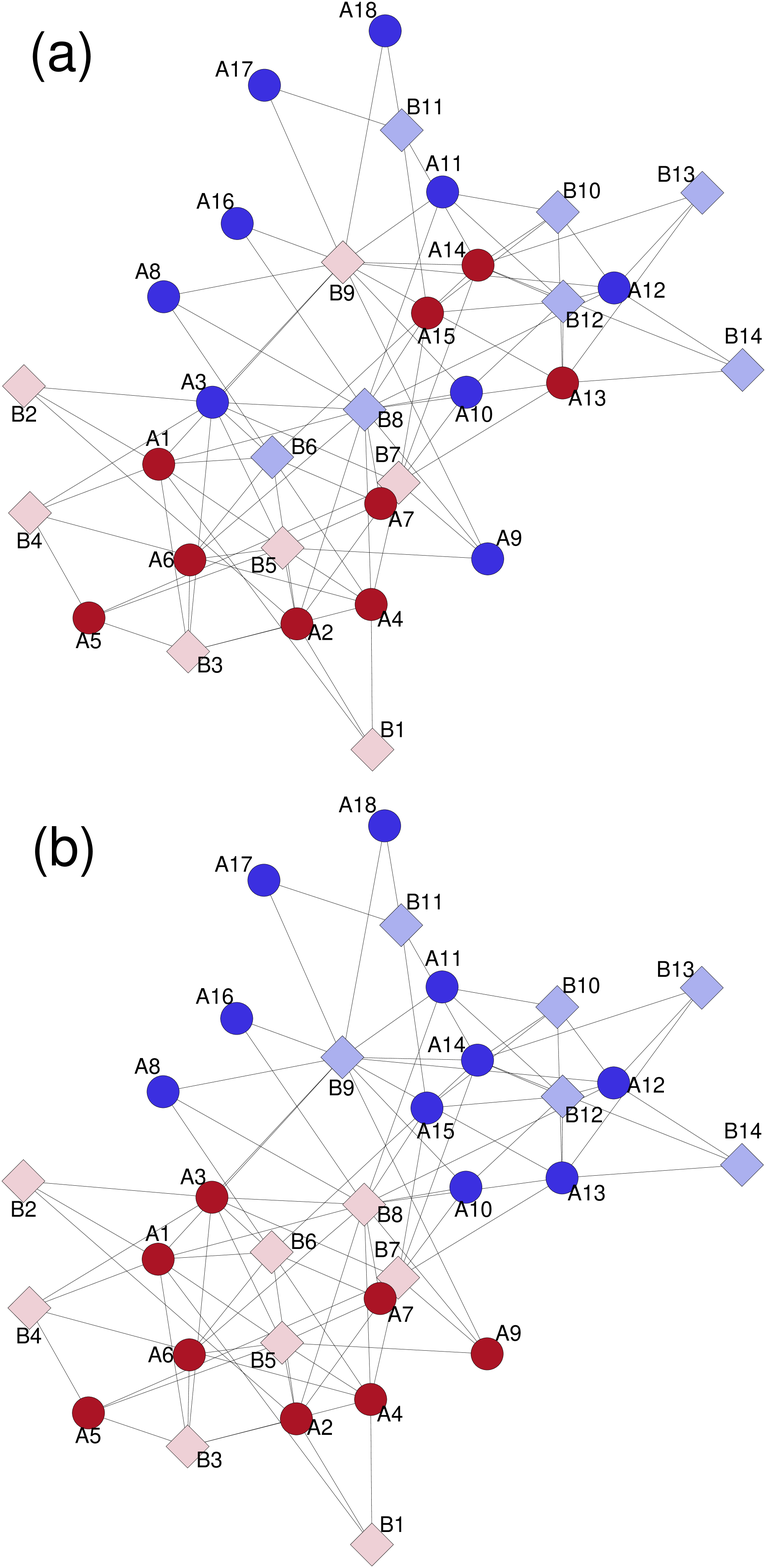} }
\caption{ Modular structure of the Southern women dataset
  \cite{davis41,freeman03}. Circles represent women and diamonds
  represent social events. A woman and an event are connected if the
  woman attended the event. (a) Modular structure as obtained
  from the unweighted projection (UWP) approach. (b) Modular
  structure as obtained from the weighted projection (WP) approach and
  the bipartite (B) approach. The UWP approach fails to capture the
  real modular structure of the network.}
\label{f-women}
\end{figure}
%
Here we analyze the modules of both women and events. We start by
considering the unweighted projection of the network in the women's
space (two women are connected if they co-attended at least one
event), and in the events' space (two events are connected if at least
one woman was in both events). As we show in Fig.~\ref{f-women}(a),
the unweighted projection does not capture the true modular structure
of the network. The failure of this approach is due to the fact that
the projections are very dense. For example, some central events were
attended by most women and thus most pairs of women are connected in
the projection.

As we show in Fig.~\ref{f-women}(b), the weighted projection approach
and the bipartite approach yield the exact same results, which do
capture the two-module structure of the network. Except for one woman,
the partition coincides with the original subjective partition
proposed by the ethnographers who collected the data, and is in
perfect agreement with some of the {\it supervised} algorithms
reviewed in Ref. \cite{freeman03}.

\section{Modules in directed networks}
\label{s-directed}

Another important class of networks for which no satisfactory module
identification algorithm has so far been proposed is directed
unipartite networks. In order to tackle this class of networks, we
note that directed networks can be conveniently represented as
bipartite networks where each node $i$ is represented by two nodes
$A_i$ and $B_i$. A directed link from $i$ to $j$ would be represented
in the bipartite network as an edge connecting $A_i$ to $B_j$.

Consider, for example, a network in which nodes are companies and
links represent investments of one company into another. By
considering each company as two different objects, one that makes
investments and one that receives investments, the directed network
can be represented as an undirected bipartite network. Modules in the
set of objects that make investments correspond to groups of companies
that invest in the same set of companies, that is, groups of companies
with a similar investing strategy.

The most widely used approach to identify communities in directed
networks is to simply disregard the directionality of the links and
identify modules using a method suitable for undirected unipartite
networks. This method might work in some situations, but will fail
when different modules are defined based on incoming and outgoing
links.

\begin{figure}[t!]
  \centerline{
    \includegraphics*[width=\columnwidth]
		     {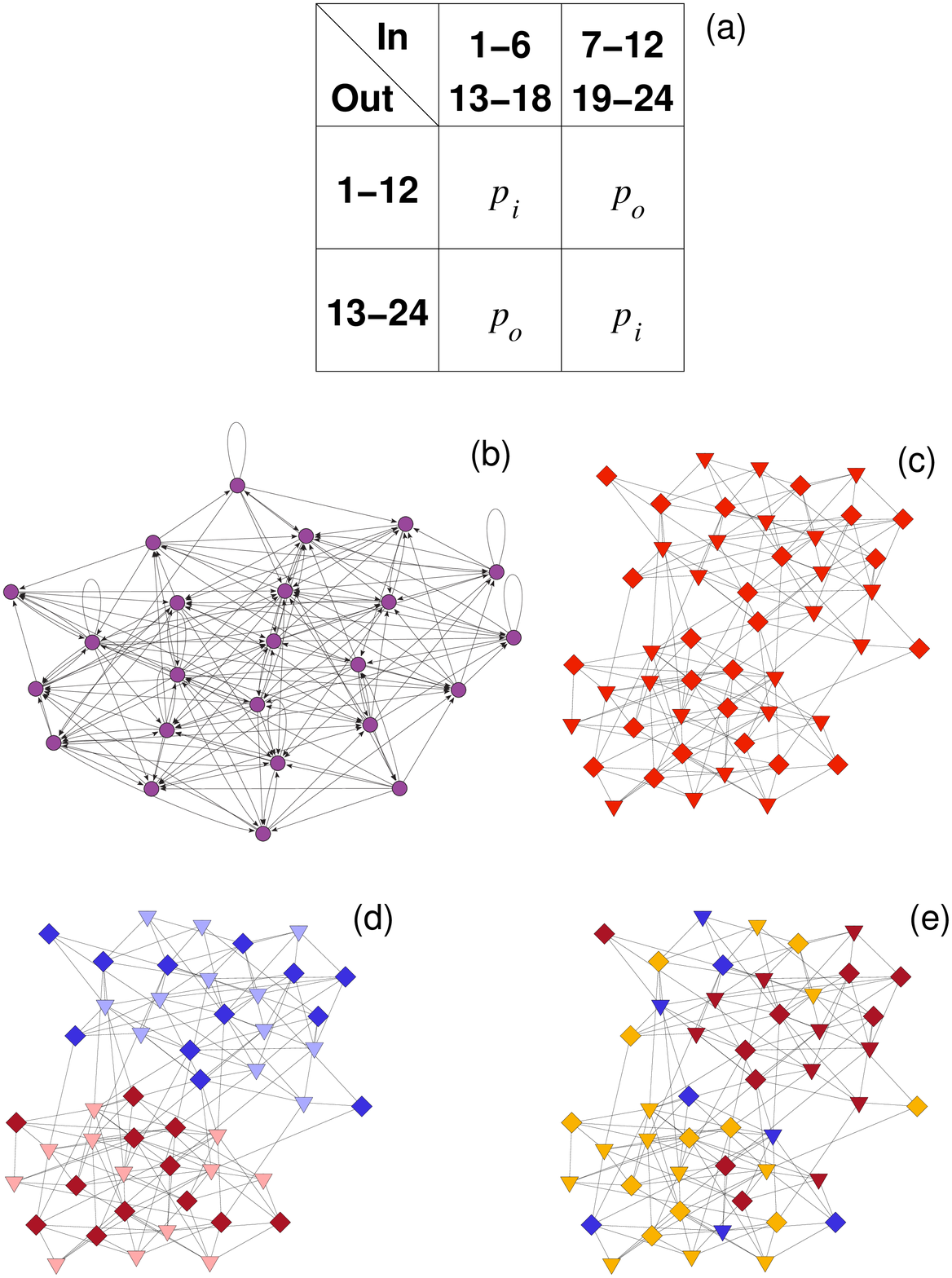}
  }
  \caption{ Application of the bipartite approach to the
  identification of modules in directed networks. (a, b) A directed
  model network. A link from node $i$ to node $j$ is established
  according to the probabilities in the matrix in (a). For example,
  there is a probability $p_i$ that there is a link from node 1 to
  node 13. In particular, we use $p_i=0.45>p_o=0.05$ to generate the
  directed network in (b). (c) Bipartite representation of the network
  in (b). Each node $i$ is in (b) is represented by two nodes here, a
  circle $A_i$ and a square $B_i$. All links in the bipartite network
  run between circles and diamonds, and a link between $A_i$ and $B_j$
  corresponds to a link from $i$ to $j$ in the directed network. (d)
  Modules identified in the bipartite network. (e) Modules identified
  from the directed network disregarding link direction. Here, we use
  the same color for $A_i$ and $B_i$, since this approach does not
  make distinctions between incoming and outgoing links.}
  \label{f-directed}
\end{figure}
%
Consider, for instance, the simple model network depicted in
Figs.~\ref{f-directed}(a, b). According to the outgoing links of the
nodes this network has two modules: nodes 1-12 and nodes
13-24. According to the incoming links of the nodes the network has
also two modules, but they are different: nodes 1-6 and 13-18 on the
one hand, and nodes 7-12 and 19-24 on the other. As we show in
Fig.~\ref{f-directed}(c), a layout of the corresponding bipartite
network already makes clear the modular structure of the network, and
any of the approaches described above (UWP, WP, and B) is able to
identify the in-modules and out-modules correctly;
Fig.~\ref{f-directed}(d). Disregarding the direction of the links,
however, results in modules that fail to capture the modular structure
of the network; Fig.~\ref{f-directed}(e).

\section{Discussion}

In this work, we have focused on approaches that aim at identifying
modules in each of the two sets of nodes in the bipartite network
independently. There are two main reasons for this choice. First,
methodologically our choice enables comparison with projection-based
algorithms, which, by definition, cannot identify modules of actors
and teams simultaneously. Second, in most situations it is reasonable
to assume that two actors belong to the same module if they
co-participate in many teams, regardless of whether the teams
themselves belong to the same module or not. An alternative approach,
however, would be to group nodes in both sets at the same time.
%

Another interesting observation relates to the optimization algorithm
used to maximize the modularity. Although we have chosen to use
simulated annealing to obtain the best possible
accuracy \cite{guimera05a,guimera05,danon05}, one can trivially use
the new modularity introduced in Eq.~(\ref{e-modularityb}) with faster
algorithms such as greedy search \cite{newman04a} or extremal
optimization \cite{duch05}.

Interestingly, one can also use the spectral methods introduced in
\cite{newman06,newman06b}. Indeed, just as the unipartite modularity
$\mathcal{M}(P)$, the bipartite modularity $\mathcal{M_B(P)}$ can be
rewritten in matrix form as
\begin{equation}
  \mathcal{M_B(P)}= \mathbf{g^T B g}\,,
\end{equation}
where $g_{is}=1$ if node $i$ belongs to module $s$ and 0 otherwise,
and the elements of the modularity matrix $\mathbf{B}$ are defined as
\begin{equation}
  B_{ij}=\left\{
  \begin{array}{ll}
    \frac{c_{ij}}{\sum_a m_a(m_a -1)} - \frac{t_i t_j}{\left( \sum_a
      m_a \right)^2} & i \neq j \\ 0 & i=j
  \end{array}\right..
\end{equation}

Even more importantly, by sampling all local maxima of the modularity
in Eq.~(\ref{e-modularityb}) one can study, not only the most modular
partition of the network, but the hierarchical structure of nested
modules and submodules \cite{sales-pardo??} within each set of nodes
in the bipartite network. This is particularly relevant taking into
account that the most modular partition of a network may, in some
cases, not represent the most ``relevant'' division of its nodes
\cite{fortunato07,sales-pardo??}.

Finally, a few words are necessary on the comparison between the
different approaches. First, we have shown that the (so far
``default'') unweighted projection approach is not reliable and can
lead, in most situations, to incorrect results. Therefore, we believe
that this approach should not be used. As for the weighted projection
approach and the bipartite approach, we have shown that their
performance is very similar, and that they are actually equivalent
when all teams in the bipartite network have the same size. We have
also pointed out, however, that they can and do give noticeably
different results when team sizes are not uniform. Given this, we
believe that the bipartite approach has a more straightforward
interpretation and would be preferable in cases in which the modular
structure of the network is unknown.


\begin{acknowledgments}
We thank R.D. Malmgren, E.N. Sawardecker, S.M.D. Seaver,
D.B. Stouffer, M.J. Stringer, and especially M.E.J. Newman and
E.A. Leicht for useful comments and suggestions. L.A.N.A. gratefully
acknowledges the support of a NIH/NIGMS K-25 award, of NSF award SBE
0624318, of the J.S. McDonnell Foundation, and of the W.~M. Keck
Foundation.
\end{acknowledgments}

\appendix
\section{Weighted unipartite modularity and bipartite modularity for bipartite networks with uniform teams}

Next, we demonstrate that, when all teams in a bipartite network have
the same size $m$, the bipartite modularity is equivalent to the
modularity of the weighted projection.

We consider the usual weighted projection, in which each pair of nodes
$i \neq j$ is connected by a link whose weight $w_{ij}$ equals the
number of times that $i$ and $j$ are together in a team; using our
previous notation $w_{ij} = c_{ij}$. No self-links are included in the
projection. 

In this projection, and when all teams have the same number of actors
$m_a \equiv m$, the constant team-size factors in
Eq.~(\ref{e-modularityb}) become
\begin{eqnarray}
  \sum_a m_a (m_a - 1) & = & N_T m (m - 1) = 2W \\
  \left( \sum_a m_a \right)^2 & = & \left( \frac{2W}{m-1} \right)^2\,
\end{eqnarray}
where, as before, $W = \sum_{i \ge j} w_{ij}$.

Each time an actor is in a team, the total weight of the links in the
projected network increases by $(m-1)$. Using this and the identities
above, we obtain
\begin{eqnarray}
  \frac{\sum_{i \ne j \in s} c_{ij}}{\sum_a m_a (m_a - 1)} & = &
      \frac{w^{\rm int}_s}{W} \\%
      \nonumber \frac{\sum_{i \ne j \in s} t_i t_j}{\left( \sum_a m_a
      \right)^2} & = & \sum_{i \in s} \frac{(m-1) t_i}{2W} \sum_{j \in
      s, j \ne i} \frac{(m-1) t_j}{2W} \\%
      & = & \left( \frac{w_s^{\rm all}}{2W} \right)^2 - \\%
      &  & \sum_{i \in s} \left( \frac{(m-1) t_i}{2W} \right)^2 \,.
\end{eqnarray}

Once the summation over modules is carried out, the last term is
simply a constant independent of the partition, and is therefore
irrelevant. Thus, up to an irrelevant constant, when all teams in a
bipartite network have the same size, the bipartite modularity in
Eq.~(\ref{e-modularityb}) is equivalent to the weighted modularity in
Eq.~(\ref{e-modularityw}).

%
%

\begin{thebibliography}{47}
\expandafter\ifx\csname natexlab\endcsname\relax\def\natexlab#1{#1}\fi
\expandafter\ifx\csname bibnamefont\endcsname\relax
  \def\bibnamefont#1{#1}\fi
\expandafter\ifx\csname bibfnamefont\endcsname\relax
  \def\bibfnamefont#1{#1}\fi
\expandafter\ifx\csname citenamefont\endcsname\relax
  \def\citenamefont#1{#1}\fi
\expandafter\ifx\csname url\endcsname\relax
  \def\url#1{\texttt{#1}}\fi
\expandafter\ifx\csname urlprefix\endcsname\relax\def\urlprefix{URL }\fi
\providecommand{\bibinfo}[2]{#2}
\providecommand{\eprint}[2][]{\url{#2}}

\bibitem[{\citenamefont{Albert and Barab\'asi}(2002)}]{albert02}
\bibinfo{author}{\bibfnamefont{R.}~\bibnamefont{Albert}} \bibnamefont{and}
  \bibinfo{author}{\bibfnamefont{A.-L.} \bibnamefont{Barab\'asi}},
  \bibinfo{journal}{Rev. Mod. Phys.} \textbf{\bibinfo{volume}{74}},
  \bibinfo{pages}{47} (\bibinfo{year}{2002}).

\bibitem[{\citenamefont{Newman}(2003)}]{newman03}
\bibinfo{author}{\bibfnamefont{M.~E.~J.} \bibnamefont{Newman}},
  \bibinfo{journal}{SIAM Review} \textbf{\bibinfo{volume}{45}},
  \bibinfo{pages}{167} (\bibinfo{year}{2003}).

\bibitem[{\citenamefont{Amaral and Ottino}(2004)}]{amaral04}
\bibinfo{author}{\bibfnamefont{L.~A.~N.} \bibnamefont{Amaral}}
  \bibnamefont{and} \bibinfo{author}{\bibfnamefont{J.}~\bibnamefont{Ottino}},
  \bibinfo{journal}{Eur. Phys. J. B} \textbf{\bibinfo{volume}{38}},
  \bibinfo{pages}{147} (\bibinfo{year}{2004}).

\bibitem[{\citenamefont{Girvan and Newman}(2002)}]{girvan02}
\bibinfo{author}{\bibfnamefont{M.}~\bibnamefont{Girvan}} \bibnamefont{and}
  \bibinfo{author}{\bibfnamefont{M.~E.~J.} \bibnamefont{Newman}},
  \bibinfo{journal}{Proc. Natl. Acad. Sci. USA} \textbf{\bibinfo{volume}{99}},
  \bibinfo{pages}{7821} (\bibinfo{year}{2002}).

\bibitem[{\citenamefont{Guimer\`a
  et~al.}(2005{\natexlab{a}})\citenamefont{Guimer\`a, Mossa, Turtschi, and
  Amaral}}]{guimera05b}
\bibinfo{author}{\bibfnamefont{R.}~\bibnamefont{Guimer\`a}},
  \bibinfo{author}{\bibfnamefont{S.}~\bibnamefont{Mossa}},
  \bibinfo{author}{\bibfnamefont{A.}~\bibnamefont{Turtschi}}, \bibnamefont{and}
  \bibinfo{author}{\bibfnamefont{L.~A.~N.} \bibnamefont{Amaral}},
  \bibinfo{journal}{Proc. Natl. Acad. Sci. USA} \textbf{\bibinfo{volume}{102}},
  \bibinfo{pages}{7794} (\bibinfo{year}{2005}{\natexlab{a}}).

\bibitem[{\citenamefont{Colizza
  et~al.}(2006{\natexlab{a}})\citenamefont{Colizza, Barrat, Barth\'elemy, and
  Vespignani}}]{colizza06a}
\bibinfo{author}{\bibfnamefont{V.}~\bibnamefont{Colizza}},
  \bibinfo{author}{\bibfnamefont{A.}~\bibnamefont{Barrat}},
  \bibinfo{author}{\bibfnamefont{M.}~\bibnamefont{Barth\'elemy}},
  \bibnamefont{and}
  \bibinfo{author}{\bibfnamefont{A.}~\bibnamefont{Vespignani}},
  \bibinfo{journal}{Proc. Natl. Acad. Sci. USA} \textbf{\bibinfo{volume}{103}},
  \bibinfo{pages}{2015} (\bibinfo{year}{2006}{\natexlab{a}}).

\bibitem[{\citenamefont{Pastor-Satorras and
  Vespignani}(2001)}]{pastor-satorras01a}
\bibinfo{author}{\bibfnamefont{R.}~\bibnamefont{Pastor-Satorras}}
  \bibnamefont{and}
  \bibinfo{author}{\bibfnamefont{A.}~\bibnamefont{Vespignani}},
  \bibinfo{journal}{Phys. Rev. Lett.} \textbf{\bibinfo{volume}{86}},
  \bibinfo{pages}{3200} (\bibinfo{year}{2001}).

\bibitem[{\citenamefont{Liljeros et~al.}(2003)\citenamefont{Liljeros, Edling,
  and Amaral}}]{liljeros03}
\bibinfo{author}{\bibfnamefont{F.}~\bibnamefont{Liljeros}},
  \bibinfo{author}{\bibfnamefont{C.~R.} \bibnamefont{Edling}},
  \bibnamefont{and} \bibinfo{author}{\bibfnamefont{L.~A.~N.}
  \bibnamefont{Amaral}}, \bibinfo{journal}{Microbes Infect.}
  \textbf{\bibinfo{volume}{5}}, \bibinfo{pages}{189} (\bibinfo{year}{2003}).

\bibitem[{\citenamefont{Guimer\`a et~al.}(2007)\citenamefont{Guimer\`a,
  Sales-Pardo, and Amaral}}]{guimera07}
\bibinfo{author}{\bibfnamefont{R.}~\bibnamefont{Guimer\`a}},
  \bibinfo{author}{\bibfnamefont{M.}~\bibnamefont{Sales-Pardo}},
  \bibnamefont{and} \bibinfo{author}{\bibfnamefont{L.~A.~N.}
  \bibnamefont{Amaral}}, \bibinfo{journal}{Nature Phys.}
  \textbf{\bibinfo{volume}{3}}, \bibinfo{pages}{63} (\bibinfo{year}{2007}).

\bibitem[{\citenamefont{Newman}(2002)}]{newman02}
\bibinfo{author}{\bibfnamefont{M.~E.~J.} \bibnamefont{Newman}},
  \bibinfo{journal}{Phys. Rev. Lett.} \textbf{\bibinfo{volume}{89}},
  \bibinfo{pages}{art. no. 208701} (\bibinfo{year}{2002}).

\bibitem[{\citenamefont{Pastor-Satorras
  et~al.}(2001)\citenamefont{Pastor-Satorras, V\'azquez, and
  Vespignani}}]{pastor-satorras01b}
\bibinfo{author}{\bibfnamefont{R.}~\bibnamefont{Pastor-Satorras}},
  \bibinfo{author}{\bibfnamefont{A.}~\bibnamefont{V\'azquez}},
  \bibnamefont{and}
  \bibinfo{author}{\bibfnamefont{A.}~\bibnamefont{Vespignani}},
  \bibinfo{journal}{Phys. Rev. Lett.} \textbf{\bibinfo{volume}{87}},
  \bibinfo{pages}{art. no. 258701} (\bibinfo{year}{2001}).

\bibitem[{\citenamefont{Maslov and Sneppen}(2002)}]{maslov02}
\bibinfo{author}{\bibfnamefont{S.}~\bibnamefont{Maslov}} \bibnamefont{and}
  \bibinfo{author}{\bibfnamefont{K.}~\bibnamefont{Sneppen}},
  \bibinfo{journal}{Science} \textbf{\bibinfo{volume}{296}},
  \bibinfo{pages}{910} (\bibinfo{year}{2002}).

\bibitem[{\citenamefont{Maslov et~al.}(2004)\citenamefont{Maslov, Sneppen, and
  Zaliznyak}}]{maslov04}
\bibinfo{author}{\bibfnamefont{S.}~\bibnamefont{Maslov}},
  \bibinfo{author}{\bibfnamefont{K.}~\bibnamefont{Sneppen}}, \bibnamefont{and}
  \bibinfo{author}{\bibfnamefont{A.}~\bibnamefont{Zaliznyak}},
  \bibinfo{journal}{Physica A} \textbf{\bibinfo{volume}{333}},
  \bibinfo{pages}{529} (\bibinfo{year}{2004}).

\bibitem[{\citenamefont{Colizza
  et~al.}(2006{\natexlab{b}})\citenamefont{Colizza, Flammini, Serrano, and
  Vespignani}}]{colizza06}
\bibinfo{author}{\bibfnamefont{V.}~\bibnamefont{Colizza}},
  \bibinfo{author}{\bibfnamefont{A.}~\bibnamefont{Flammini}},
  \bibinfo{author}{\bibfnamefont{M.~A.} \bibnamefont{Serrano}},
  \bibnamefont{and}
  \bibinfo{author}{\bibfnamefont{A.}~\bibnamefont{Vespignani}},
  \bibinfo{journal}{Nature Phys.} \textbf{\bibinfo{volume}{2}},
  \bibinfo{pages}{110} (\bibinfo{year}{2006}{\natexlab{b}}).

\bibitem[{\citenamefont{Danon et~al.}(2005)\citenamefont{Danon,
  D\'{\i}az-Guilera, Duch, and Arenas}}]{danon05}
\bibinfo{author}{\bibfnamefont{L.}~\bibnamefont{Danon}},
  \bibinfo{author}{\bibfnamefont{A.}~\bibnamefont{D\'{\i}az-Guilera}},
  \bibinfo{author}{\bibfnamefont{J.}~\bibnamefont{Duch}}, \bibnamefont{and}
  \bibinfo{author}{\bibfnamefont{A.}~\bibnamefont{Arenas}},
  \bibinfo{journal}{J. Stat. Mech.: Theor. Exp.} p. \bibinfo{pages}{art. no.
  P09008} (\bibinfo{year}{2005}).

\bibitem[{\citenamefont{Borgatti and Everett}(1997)}]{borgatti97}
\bibinfo{author}{\bibfnamefont{S.~P.} \bibnamefont{Borgatti}} \bibnamefont{and}
  \bibinfo{author}{\bibfnamefont{M.~G.} \bibnamefont{Everett}},
  \bibinfo{journal}{Social Networks} \textbf{\bibinfo{volume}{19}},
  \bibinfo{pages}{243} (\bibinfo{year}{1997}).

\bibitem[{\citenamefont{Doreian et~al.}(2004)\citenamefont{Doreian, Batagelj,
  and Ferligoj}}]{doreian04}
\bibinfo{author}{\bibfnamefont{P.}~\bibnamefont{Doreian}},
  \bibinfo{author}{\bibfnamefont{V.}~\bibnamefont{Batagelj}}, \bibnamefont{and}
  \bibinfo{author}{\bibfnamefont{A.}~\bibnamefont{Ferligoj}},
  \bibinfo{journal}{Social Networks} \textbf{\bibinfo{volume}{26}},
  \bibinfo{pages}{29} (\bibinfo{year}{2004}).

\bibitem[{\citenamefont{Uetz et~al.}(2000)\citenamefont{Uetz, Giot, Cagney,
  Mansfield, Judson, Knight, Lockshon, Narayan, Srinivasan, Pochart
  et~al.}}]{uetz00}
\bibinfo{author}{\bibfnamefont{P.}~\bibnamefont{Uetz}},
  \bibinfo{author}{\bibfnamefont{L.}~\bibnamefont{Giot}},
  \bibinfo{author}{\bibfnamefont{G.}~\bibnamefont{Cagney}},
  \bibinfo{author}{\bibfnamefont{T.~A.} \bibnamefont{Mansfield}},
  \bibinfo{author}{\bibfnamefont{R.~S.} \bibnamefont{Judson}},
  \bibinfo{author}{\bibfnamefont{J.~R.} \bibnamefont{Knight}},
  \bibinfo{author}{\bibfnamefont{D.}~\bibnamefont{Lockshon}},
  \bibinfo{author}{\bibfnamefont{V.}~\bibnamefont{Narayan}},
  \bibinfo{author}{\bibfnamefont{M.}~\bibnamefont{Srinivasan}},
  \bibinfo{author}{\bibfnamefont{P.}~\bibnamefont{Pochart}},
  \bibnamefont{et~al.}, \bibinfo{journal}{Nature}
  \textbf{\bibinfo{volume}{403}}, \bibinfo{pages}{623} (\bibinfo{year}{2000}).

\bibitem[{\citenamefont{Jeong et~al.}(2001)\citenamefont{Jeong, Mason,
  Barab\'asi, and Oltvai}}]{jeong01}
\bibinfo{author}{\bibfnamefont{H.}~\bibnamefont{Jeong}},
  \bibinfo{author}{\bibfnamefont{S.~P.} \bibnamefont{Mason}},
  \bibinfo{author}{\bibfnamefont{A.-L.} \bibnamefont{Barab\'asi}},
  \bibnamefont{and} \bibinfo{author}{\bibfnamefont{Z.~N.}
  \bibnamefont{Oltvai}}, \bibinfo{journal}{Nature}
  \textbf{\bibinfo{volume}{411}}, \bibinfo{pages}{41} (\bibinfo{year}{2001}).

\bibitem[{\citenamefont{Li et~al.}(2004)\citenamefont{Li, Armstrong, Bertin,
  Ge, Milstein, Boxem, Vidalain, Han, Chesneau, Hao et~al.}}]{li04}
\bibinfo{author}{\bibfnamefont{S.}~\bibnamefont{Li}},
  \bibinfo{author}{\bibfnamefont{C.~M.} \bibnamefont{Armstrong}},
  \bibinfo{author}{\bibfnamefont{N.}~\bibnamefont{Bertin}},
  \bibinfo{author}{\bibfnamefont{H.}~\bibnamefont{Ge}},
  \bibinfo{author}{\bibfnamefont{S.}~\bibnamefont{Milstein}},
  \bibinfo{author}{\bibfnamefont{M.}~\bibnamefont{Boxem}},
  \bibinfo{author}{\bibfnamefont{P.-O.} \bibnamefont{Vidalain}},
  \bibinfo{author}{\bibfnamefont{J.-D.~J.} \bibnamefont{Han}},
  \bibinfo{author}{\bibfnamefont{A.}~\bibnamefont{Chesneau}},
  \bibinfo{author}{\bibfnamefont{T.}~\bibnamefont{Hao}}, \bibnamefont{et~al.},
  \bibinfo{journal}{Science} \textbf{\bibinfo{volume}{303}},
  \bibinfo{pages}{540} (\bibinfo{year}{2004}).

\bibitem[{\citenamefont{Jordano}(1987)}]{jordano87}
\bibinfo{author}{\bibfnamefont{P.}~\bibnamefont{Jordano}},
  \bibinfo{journal}{Am. Nat.} \textbf{\bibinfo{volume}{129}},
  \bibinfo{pages}{657} (\bibinfo{year}{1987}).

\bibitem[{\citenamefont{Bascompte et~al.}(2003)\citenamefont{Bascompte,
  Jordano, Melián, and Olesen}}]{bascompte03}
\bibinfo{author}{\bibfnamefont{J.}~\bibnamefont{Bascompte}},
  \bibinfo{author}{\bibfnamefont{P.}~\bibnamefont{Jordano}},
  \bibinfo{author}{\bibfnamefont{C.~J.} \bibnamefont{Melián}},
  \bibnamefont{and} \bibinfo{author}{\bibfnamefont{J.~M.}
  \bibnamefont{Olesen}}, \bibinfo{journal}{Proc. Natl. Acad. Sci. U. S. A.}
  \textbf{\bibinfo{volume}{100}}, \bibinfo{pages}{9383} (\bibinfo{year}{2003}).

\bibitem[{\citenamefont{Newman}(2001)}]{newman01b}
\bibinfo{author}{\bibfnamefont{M.~E.~J.} \bibnamefont{Newman}},
  \bibinfo{journal}{Proc. Natl. Acad. Sci. USA} \textbf{\bibinfo{volume}{98}},
  \bibinfo{pages}{404} (\bibinfo{year}{2001}).

\bibitem[{\citenamefont{B\"orner et~al.}(2004)\citenamefont{B\"orner, Maru, and
  Goldstone}}]{borner04}
\bibinfo{author}{\bibfnamefont{K.}~\bibnamefont{B\"orner}},
  \bibinfo{author}{\bibfnamefont{J.~T.} \bibnamefont{Maru}}, \bibnamefont{and}
  \bibinfo{author}{\bibfnamefont{R.~L.} \bibnamefont{Goldstone}},
  \bibinfo{journal}{Proc. Natl. Acad. Sci. USA} \textbf{\bibinfo{volume}{101}},
  \bibinfo{pages}{5266} (\bibinfo{year}{2004}).

\bibitem[{\citenamefont{Guimer\`a
  et~al.}(2005{\natexlab{b}})\citenamefont{Guimer\`a, Uzzi, Spiro, and
  Amaral}}]{guimera05c}
\bibinfo{author}{\bibfnamefont{R.}~\bibnamefont{Guimer\`a}},
  \bibinfo{author}{\bibfnamefont{B.}~\bibnamefont{Uzzi}},
  \bibinfo{author}{\bibfnamefont{J.}~\bibnamefont{Spiro}}, \bibnamefont{and}
  \bibinfo{author}{\bibfnamefont{L.~A.~N.} \bibnamefont{Amaral}},
  \bibinfo{journal}{Science} \textbf{\bibinfo{volume}{308}},
  \bibinfo{pages}{697} (\bibinfo{year}{2005}{\natexlab{b}}).

\bibitem[{\citenamefont{Gleiser and Danon}(2003)}]{gleiser03}
\bibinfo{author}{\bibfnamefont{P.~M.} \bibnamefont{Gleiser}} \bibnamefont{and}
  \bibinfo{author}{\bibfnamefont{L.}~\bibnamefont{Danon}},
  \bibinfo{journal}{Adv. Complex Syst.} \textbf{\bibinfo{volume}{6}},
  \bibinfo{pages}{565} (\bibinfo{year}{2003}).

\bibitem[{\citenamefont{Uzzi and Spiro}(2005)}]{uzzi05}
\bibinfo{author}{\bibfnamefont{B.}~\bibnamefont{Uzzi}} \bibnamefont{and}
  \bibinfo{author}{\bibfnamefont{J.}~\bibnamefont{Spiro}},
  \bibinfo{journal}{Am. J. Sociol.} \textbf{\bibinfo{volume}{111}},
  \bibinfo{pages}{447} (\bibinfo{year}{2005}).

\bibitem[{\citenamefont{Stouffer et~al.}(2005)\citenamefont{Stouffer, Camacho,
  Guimer\`a, Ng, and Amaral}}]{stouffer05}
\bibinfo{author}{\bibfnamefont{D.~B.} \bibnamefont{Stouffer}},
  \bibinfo{author}{\bibfnamefont{J.}~\bibnamefont{Camacho}},
  \bibinfo{author}{\bibfnamefont{R.}~\bibnamefont{Guimer\`a}},
  \bibinfo{author}{\bibfnamefont{C.~A.} \bibnamefont{Ng}}, \bibnamefont{and}
  \bibinfo{author}{\bibfnamefont{L.~A.~N.} \bibnamefont{Amaral}},
  \bibinfo{journal}{Ecology} \textbf{\bibinfo{volume}{86}},
  \bibinfo{pages}{1301} (\bibinfo{year}{2005}).

\bibitem[{\citenamefont{Williams and Martinez}(2000)}]{williams00}
\bibinfo{author}{\bibfnamefont{R.~J.} \bibnamefont{Williams}} \bibnamefont{and}
  \bibinfo{author}{\bibfnamefont{N.~D.} \bibnamefont{Martinez}},
  \bibinfo{journal}{Nature} \textbf{\bibinfo{volume}{404}},
  \bibinfo{pages}{180} (\bibinfo{year}{2000}).

\bibitem[{\citenamefont{Barab\'asi and Oltvai}(2004)}]{barabasi04}
\bibinfo{author}{\bibfnamefont{A.-L.} \bibnamefont{Barab\'asi}}
  \bibnamefont{and} \bibinfo{author}{\bibfnamefont{Z.~N.}
  \bibnamefont{Oltvai}}, \bibinfo{journal}{Nat. Rev. Genet.}
  \textbf{\bibinfo{volume}{5}}, \bibinfo{pages}{101} (\bibinfo{year}{2004}).

\bibitem[{\citenamefont{Everitt et~al.}(2001)\citenamefont{Everitt, Landau, and
  Leese}}]{everitt01}
\bibinfo{author}{\bibfnamefont{B.~S.} \bibnamefont{Everitt}},
  \bibinfo{author}{\bibfnamefont{S.}~\bibnamefont{Landau}}, \bibnamefont{and}
  \bibinfo{author}{\bibfnamefont{M.}~\bibnamefont{Leese}},
  \emph{\bibinfo{title}{Cluster Analysis}} (\bibinfo{publisher}{Arnold Pub.},
  \bibinfo{year}{2001}).

\bibitem[{\citenamefont{Freenan}(2003)}]{freeman03}
\bibinfo{author}{\bibfnamefont{L.~C.} \bibnamefont{Freenan}}, in
  \emph{\bibinfo{booktitle}{Dynamic Social Network Modeling and Analysis:
  Workshop Summary and Papers}}, edited by
  \bibinfo{editor}{\bibfnamefont{R.}~\bibnamefont{Breiger}},
  \bibinfo{editor}{\bibfnamefont{C.}~\bibnamefont{Carley}}, \bibnamefont{and}
  \bibinfo{editor}{\bibfnamefont{P.}~\bibnamefont{Pattison}}
  (\bibinfo{publisher}{The National Academies Press},
  \bibinfo{address}{Washington, DC}, \bibinfo{year}{2003}), pp.
  \bibinfo{pages}{39--97}.

\bibitem[{\citenamefont{Fortunato and Barth\'elemy}(2007)}]{fortunato07}
\bibinfo{author}{\bibfnamefont{S.}~\bibnamefont{Fortunato}} \bibnamefont{and}
  \bibinfo{author}{\bibfnamefont{M.}~\bibnamefont{Barth\'elemy}},
  \bibinfo{journal}{Proc. Natl. Acad. Sci. USA} \textbf{\bibinfo{volume}{104}},
  \bibinfo{pages}{36} (\bibinfo{year}{2007}).

\bibitem[{\citenamefont{Sales-Pardo et~al.}(2007)\citenamefont{Sales-Pardo,
  Guimer\`a, Moreira, and Amaral}}]{sales-pardo??}
\bibinfo{author}{\bibfnamefont{M.}~\bibnamefont{Sales-Pardo}},
  \bibinfo{author}{\bibfnamefont{R.}~\bibnamefont{Guimer\`a}},
  \bibinfo{author}{\bibfnamefont{A.~A.} \bibnamefont{Moreira}},
  \bibnamefont{and} \bibinfo{author}{\bibfnamefont{L.~A.~N.}
  \bibnamefont{Amaral}}, \bibinfo{journal}{arXiv:0705.1679}
  (\bibinfo{year}{2007}).

\bibitem[{\citenamefont{Fortunato}(2007)}]{fortunato07a}
\bibinfo{author}{\bibfnamefont{S.}~\bibnamefont{Fortunato}},
  \bibinfo{journal}{arXiv:0705.4445}  (\bibinfo{year}{2007}).

\bibitem[{\citenamefont{Guimer\`a and Amaral}(2005{\natexlab{a}})}]{guimera05a}
\bibinfo{author}{\bibfnamefont{R.}~\bibnamefont{Guimer\`a}} \bibnamefont{and}
  \bibinfo{author}{\bibfnamefont{L.~A.~N.} \bibnamefont{Amaral}},
  \bibinfo{journal}{Nature} \textbf{\bibinfo{volume}{433}},
  \bibinfo{pages}{895} (\bibinfo{year}{2005}{\natexlab{a}}).

\bibitem[{\citenamefont{Guimer\`a and Amaral}(2005{\natexlab{b}})}]{guimera05}
\bibinfo{author}{\bibfnamefont{R.}~\bibnamefont{Guimer\`a}} \bibnamefont{and}
  \bibinfo{author}{\bibfnamefont{L.~A.~N.} \bibnamefont{Amaral}},
  \bibinfo{journal}{J. Stat. Mech.: Theor. Exp.} p. \bibinfo{pages}{art. no.
  P02001} (\bibinfo{year}{2005}{\natexlab{b}}).

\bibitem[{\citenamefont{Guimer\`a et~al.}(2004)\citenamefont{Guimer\`a,
  Sales-Pardo, and Amaral}}]{guimera04}
\bibinfo{author}{\bibfnamefont{R.}~\bibnamefont{Guimer\`a}},
  \bibinfo{author}{\bibfnamefont{M.}~\bibnamefont{Sales-Pardo}},
  \bibnamefont{and} \bibinfo{author}{\bibfnamefont{L.~A.~N.}
  \bibnamefont{Amaral}}, \bibinfo{journal}{Phys. Rev. E}
  \textbf{\bibinfo{volume}{70}}, \bibinfo{pages}{art. no. 025101}
  (\bibinfo{year}{2004}).

\bibitem[{\citenamefont{Reichardt and Bornholdt}(2006)}]{reichardt06}
\bibinfo{author}{\bibfnamefont{J.}~\bibnamefont{Reichardt}} \bibnamefont{and}
  \bibinfo{author}{\bibfnamefont{S.}~\bibnamefont{Bornholdt}},
  \bibinfo{journal}{Phys. Rev. E} \textbf{\bibinfo{volume}{74}},
  \bibinfo{pages}{016110} (\bibinfo{year}{2006}).

\bibitem[{\citenamefont{Newman and Girvan}(2004)}]{newman04b}
\bibinfo{author}{\bibfnamefont{M.~E.~J.} \bibnamefont{Newman}}
  \bibnamefont{and} \bibinfo{author}{\bibfnamefont{M.}~\bibnamefont{Girvan}},
  \bibinfo{journal}{Phys. Rev. E} \textbf{\bibinfo{volume}{69}},
  \bibinfo{pages}{art. no. 026113} (\bibinfo{year}{2004}).

\bibitem[{\citenamefont{Kirkpatrick et~al.}(1983)\citenamefont{Kirkpatrick,
  Gelatt, and Vecchi}}]{kirkpatrick83}
\bibinfo{author}{\bibfnamefont{S.}~\bibnamefont{Kirkpatrick}},
  \bibinfo{author}{\bibfnamefont{C.~D.} \bibnamefont{Gelatt}},
  \bibnamefont{and} \bibinfo{author}{\bibfnamefont{M.~P.}
  \bibnamefont{Vecchi}}, \bibinfo{journal}{Science}
  \textbf{\bibinfo{volume}{220}}, \bibinfo{pages}{671} (\bibinfo{year}{1983}).

\bibitem[{\citenamefont{Newman}(2004)}]{newman04a}
\bibinfo{author}{\bibfnamefont{M.~E.~J.} \bibnamefont{Newman}},
  \bibinfo{journal}{Phys. Rev. E} \textbf{\bibinfo{volume}{69}},
  \bibinfo{pages}{art. no. 066133} (\bibinfo{year}{2004}).

\bibitem[{\citenamefont{Duch and Arenas}(2005)}]{duch05}
\bibinfo{author}{\bibfnamefont{J.}~\bibnamefont{Duch}} \bibnamefont{and}
  \bibinfo{author}{\bibfnamefont{A.}~\bibnamefont{Arenas}},
  \bibinfo{journal}{Phys. Rev. E} \textbf{\bibinfo{volume}{72}},
  \bibinfo{pages}{art. no. 027104} (\bibinfo{year}{2005}).

\bibitem[{\citenamefont{Newman}(2006{\natexlab{a}})}]{newman06}
\bibinfo{author}{\bibfnamefont{M.~E.~J.} \bibnamefont{Newman}},
  \bibinfo{journal}{Proc. Natl. Acad. Sci. USA} \textbf{\bibinfo{volume}{103}},
  \bibinfo{pages}{8577} (\bibinfo{year}{2006}{\natexlab{a}}).

\bibitem[{\citenamefont{Newman}(2006{\natexlab{b}})}]{newman06b}
\bibinfo{author}{\bibfnamefont{M.~E.~J.} \bibnamefont{Newman}},
  \bibinfo{journal}{Phys. Rev. E} \textbf{\bibinfo{volume}{74}},
  \bibinfo{pages}{art. no. 036104} (\bibinfo{year}{2006}{\natexlab{b}}).

\bibitem[{\citenamefont{Danon et~al.}(2006)\citenamefont{Danon,
  D\'{\i}az-Guilera, and Arenas}}]{danon06}
\bibinfo{author}{\bibfnamefont{L.}~\bibnamefont{Danon}},
  \bibinfo{author}{\bibfnamefont{A.}~\bibnamefont{D\'{\i}az-Guilera}},
  \bibnamefont{and} \bibinfo{author}{\bibfnamefont{A.}~\bibnamefont{Arenas}},
  \bibinfo{journal}{J. Stat. Mech.: Theor. Exp.} p. \bibinfo{pages}{P11010}
  (\bibinfo{year}{2006}).

\bibitem[{\citenamefont{Davis et~al.}(1941)\citenamefont{Davis, Gardner, and
  Gardner}}]{davis41}
\bibinfo{author}{\bibfnamefont{A.}~\bibnamefont{Davis}},
  \bibinfo{author}{\bibfnamefont{B.~B.} \bibnamefont{Gardner}},
  \bibnamefont{and} \bibinfo{author}{\bibfnamefont{M.~R.}
  \bibnamefont{Gardner}}, \emph{\bibinfo{title}{Deep {S}outh}}
  (\bibinfo{publisher}{University of Chicago Press},
  \bibinfo{address}{Chicago}, \bibinfo{year}{1941}).

\end{thebibliography}

\end{document}